\documentclass[sigconf,nonacm]{acmart}
\AtBeginDocument{%
  }

\usepackage{amsmath}
\usepackage{amsthm}
\usepackage{graphicx}
\usepackage{subcaption}
\usepackage{tikz}
\usepackage{listings}
\usepackage[english]{babel}
\newcommand{\xb}{\mathbf{x}}

\newcommand{\er}{Erd{\H{o}}s--R{\'e}nyi}

\lstdefinelanguage{makefile}{
    morekeywords={all, clean, gcc},
    sensitive=false,
    morecomment=[l]{\#},
}

\def\kb{{\mathbf k}}
\def\xb{{\mathbf x}}

\def\Sc{{\mathcal{S}}}
\def\Gk{{\mathfrak{G}}}
\def\Uk{{\mathfrak{U}}}

\def\RR{{\mathbb{R}}}

\def\ad{{\rm ad}}
\def\cd{{\rm cd}}

\def\lcm{{\rm lcm}}
\def\bd{{\rm bd}}
\def\lbd{{\rm lbd}}

\theoremstyle{definition}
\newtheorem{example}{Example}[section]

\newtheorem{remark}{Remark}[section]

\copyrightyear{2026}
\acmYear{2026}
\setcopyright{cc}
\setcctype{by-nc-nd}
\acmConference[ISSAC '26]{51st International Symposium on Symbolic and Algebraic Computation}{July 13--17, 2026}{Oldenburg, Germany}
\acmBooktitle{51st International Symposium on Symbolic and Algebraic Computation (ISSAC '26), July 13--17, 2026, Oldenburg, Germany}
\acmDOI{10.1145/3815436.3815458}
\acmISBN{979-8-4007-2595-1/2026/07}

\begin{document}

\title{Asymptotic properties of random monomial ideals}

\author{Fatemeh Mohammadi}
\email{fatemeh.mohammadi@kuleuven.be}
\orcid{0000-0001-5187-0995}
\affiliation{%
  \institution{KU Leuven}
  \city{Leuven}
  \country{Belgium}
}

\author{Sonja Petrovi\'c}
\email{sonja.petrovic@illinoistech.edu}
\orcid{0000-0002-4784-4169}
\affiliation{%
  \institution{Illinois Institute of Technology}
  \city{Chicago}
  \country{USA}
}

\author{Eduardo S\'aenz-de-Cabez\'on}
\email{esaenz-d@unirioja.es}
\orcid{0000-0002-5615-4194}
\affiliation{%
  \institution{Universidad de La Rioja}
  \city{Logro\~no}
  \country{Spain}
}


\begin{abstract}
This paper focuses on asymptotic properties of random monomial ideals through a statistical viewpoint. It extends the study of redundancy in monomial ideals by analyzing the poset density of the LCM-lattice. We explore how this density behaves across random algebraic models and structured networks. Experimental data reveal that the LCM-lattice exhibits sharp threshold behavior rather than changing smoothly. We observe a strong negative correlation between the number of generators and LCM-lattice density, abruptly separating three distinct regimes: a low-density Taylor-like regime, a high-density redundant regime, and a narrow transition window. We show that increasing the generator degree causes this density drop to occur at lower probability thresholds. We conclude by conjecturing that for equigenerated squarefree ideals, the LCM-lattice density undergoes a sharp phase transition, analogous to the emergence of giant components in hypergraphs. 
This suggests that the classical, ideal-by-ideal role of the LCM-lattice as a combinatorial invariant also admits a statistical/asymptotic counterpart: in natural random families, redundancy and resolution-complexity indicators concentrate into distinct typical regimes separated by a narrow transition window. 
\end{abstract}

\ccsdesc[500]{Mathematics of computing~Combinatoric problems}
\ccsdesc[500]{Applied computing~Mathematics and statistics}

\keywords{monomial ideals, random models, asymptotic properties, lcm-lattice}



\renewcommand\footnotetextcopyrightpermission[1]{}

\acmConference[ISSAC '26]{51st International Symposium on Symbolic and Algebraic Computation}{July 13--17, 2026}{Oldenburg, Germany}

\acmDOI{10.1145/3815436.3815458}

\settopmatter{printacmref=false}

\maketitle

\begin{center}
\small
Accepted author manuscript. The final version of record appears in the
Proceedings of ISSAC '26, ACM, DOI: 10.1145/3815436.3815458.
\end{center}

\section{Introduction}

Monomial ideals form a central object of study at the interface of commutative algebra and combinatorics. Let $R=k[x_{1},...,x_{n}]$ be the polynomial ring on n variables over a field k.  
Classical work studies combinatorial invariants attached to an individual ideal, and among the most fundamental is the LCM-lattice: its size and shape encode overlap among minimal generators and control homological invariants such as multigraded Betti numbers and minimal free resolutions. In this traditional perspective, LCM-lattices are analyzed ideal-by-ideal, without a notion of ``typical" or asymptotic behavior across a family. The goal of this paper is to introduce a complementary statistical viewpoint. We study random models on structural sets of monomials and track how algebraic interactions among randomly chosen generators collapse into fewer distinct least common multiples as the sampling parameter varies. To quantify this, we use the (normalized) size of the LCM-lattice, that is, algebraic (poset) density, and we also record derived resolution-level summaries. This framework allows us to ask asymptotic questions about the LCM-lattice and related homological data across large random families. 

The motivation for this study includes the knowledge of the asymptotic behavior of theoretical properties of monomial ideals, in the vein of \cite{DPSSW19,camarneiro2022convex,silverstein2023asymptotic}, or in applications of monomial ideals to other areas, see \cite{MSW25} and references therein. 

Our approach in this paper is strictly that of experimental mathematics \cite{bailey2007experimental} in relation to random models in networks \cite{ER59} and random monomial ideals \cite{DPSSW19}. Rather than presenting a new algorithm or theoretical proofs, our primary contribution lies in structuring a large-scale computational framework to probe these systems. This experimental data reveals subtle asymptotic behaviors that classical, ideal-by-ideal analysis misses, leading to our central conjecture.

\begin{definition}
Let $\Uk$ be a set of monomials in $R$. We say that $\Uk$ is a {\em structural set of monomials} if every nonempty subset $\Sc\subseteq\Uk$
 is the minimal generating set of a monomial ideal in $R$, i.e. for every pair $\xb^\mu\neq\xb^\nu$ in $\Sc$, we have that $\xb^\mu\nmid\xb^\nu$ and  $\xb^\nu\nmid\xb^\mu$.
 
 For every set $\Sc\subseteq\mathfrak\Uk$ we denote by $I_\Sc$ the monomial ideal minimally generated by the elements of $\Sc$.
\end{definition}
Structural sets provide a setting in which every subset of generators defines
a monomial ideal without redundancies. This allows us to compare the size of a
generating set with the complexity of the algebraic interactions it induces.
We formalize this comparison via the following two notions of density.

\begin{definition}
Let $\Uk\subseteq R$ be a structural set of monomials, and $\Sc\subseteq\Uk$. The {\em combinatorial density} of $\Sc$ is given by 
\[
\cd(\Sc)=\frac{|\Sc|}{|\Uk|},
\]
and the {\em algebraic density} of $\Sc$ is given by
\[
\ad(\Sc)=\frac{|L_{I_\Sc}|}{2^{|\Sc|}},
\]
where $L_{I_\Sc}$ is the $\lcm$-lattice of $I_\Sc$. We shall also use the term {\em poset density} or {\em $\lcm$-density} for the algebraic density, as stated in \cite{MSW25}. The normalization by $2^{|\Sc|}$ corresponds to the maximal possible size
of the lcm-lattice, attained when all subsets of generators produce
distinct least common multiples. 
\end{definition}

Derived from this algebraic notion of density, we can define the following, related to the shape of the Betti diagram of the ideal, i.e. the shape of its minimal free resolution..
\begin{definition}
    Let $I\subseteq\kb[x_1\dots,x_n]$ be a monomial ideal. The \em{Betti density} of $I$, $\bd(I)$ is defined by 
    \[
     \bd(I)=\frac{\sum_{\beta_i(I)\neq0}\beta_i(I)}{2^{|\Sc|}},
    \]
and the {\em length Betti density} of $I$ is defined by
\[
     \lbd(I)=\frac{|\{j\mid \beta_{i,j}(I)\neq0\}|}{2^{|\Sc|}},
    \]
\end{definition}

The algebraic density reflects the structure of the $\lcm$-lattice, which
encodes how subsets of generators combine via least common multiples.
Distinct subsets may yield the same least common multiple, and such
coincidences reduce the size of the lattice. In this way, the $\lcm$-density
measures the extent of algebraic redundancy among the generators.
Formal definitions and basic properties of $\lcm$-lattices are recalled in
Section~\ref{sec:pre}.

\begin{definition}
A \emph{random model} on a structural monomial set $\Uk$ is a stochastic process
that randomly selects a subset $\Sc \subseteq \Uk$, and hence a monomial ideal
$I_\Sc$.
\end{definition}
In this way, randomness provides a mechanism for probing typical
interaction patterns rather than properties of individual ideals.

\medskip 
Our main goal in this paper is to study asymptotic properties of families of such
random models on structural monomial sets, with particular emphasis on the size
and structure of the associated $\lcm$-lattices and filtrations. 
In particular, we focus on the structural sets $\Uk_{(1,n,d)}$ consisting of all squarefree monomials of degree $d$ in $R=\kb[x_1,\dots,x_n]$, where the subscript $1$ indicates that the maximum exponent of any variable is 1, ensuring the monomials are squarefree.
The case $d=2$ corresponds to squarefree monomial ideals generated in degree two,
equivalently to edge ideals of simple graphs on $n$ vertices. From this
perspective, classical random graph models arise as special cases of random
monomial ideals. Among these, the Erd\H{o}s--R\'enyi model plays a central role and
serves both as a guiding example and as motivation for the general theory
developed in this paper, where algebraic phase transitions mirror classical
percolation phenomena.

\medskip
\noindent
\textbf{Main theme.}
We study phase transition phenomena in the algebraic structure of random monomial ideals.  Our experimental results suggest that when monomial ideals are sampled from natural random models, the LCM-lattice exhibits sharp phase transition behavior. Rather than changing smoothly with the number or degree of generators, the lattice size collapses abruptly at a critical probability, separating three distinct regimes.

This behavior is not predicted by classical theory. It suggests that LCM-lattices admit threshold phenomena analogous to the emergence of giant components in Erd\H{o}s--R\'enyi random graphs and their higher-dimensional analogues.  From this perspective, the conjectured phase transition reframes classical notions such as genericity, redundancy, and Taylor behavior as emergent regimes in large random systems, providing new insight into how algebraic complexity typically arises—or collapses—in families of monomial ideals.

\noindent\textbf{Asymptotic regimes for a classical invariant.} 
Our experiments indicate that, in these random families, the LCM-lattice does not vary smoothly with the number or degree of generators:
its normalized size exhibits a sharp collapse at a critical probability, separating three regimes (Taylor-like / low redundancy, highly redundant, and a narrow transition window). Because the LCM-lattice governs multidegrees and supports of the minimal free resolution, this threshold behavior provides an asymptotic organization principle for standard commutative-algebra features: it predicts parameter ranges in which resolution-level summaries (such as the distribution of nonzero Betti numbers and effective resolution length, as captured by our Betti-density statistics) are typically stable, and a narrow range in which they change rapidly. The observed systematic dependence of the threshold location on generator degree highlights the role of higher-order overlaps (not generator count) influencing when redundancy becomes typical in large families.

\medskip
\noindent\textbf{Outline.}
Section~\ref{sec:pre} recalls the $\lcm$-lattice, which provides an algebraic framework for measuring interactions among generators of monomial ideals. Section~\ref{sec:ER} reviews the Erd{\H{o}}s--R{\'e}nyi random graph model and its classical phase transition behavior. Section~\ref{sec:algER} develops an algebraic interpretation of these phenomena for equigenerated monomial ideals, revealing phase-transition behavior analogous to classical random graph models. In Subsection~\ref{sec:algER1}, we study this behavior for edge ideals of graphs by identifying a phase transition in the $\operatorname{lcm}$-density. Subsection~\ref{sec:algER2} extends this analysis to equigenerated squarefree monomial ideals of higher degree, corresponding to uniform hypergraphs, and presents empirical evidence for analogous behavior.

\section{The $\lcm$-lattice}\label{sec:pre}

The interactions among the minimal generators of a monomial ideal are encoded
by their least common multiples. These interactions are organized in a
combinatorial object called the $\lcm$-lattice, which plays a central role in
the homological and combinatorial study of monomial ideals.

Let $M = \{\xb^{\mu_1}, \dots, \xb^{\mu_r}\}$ be a finite set of monomials in
$R=\kb[x_1,\dots,x_n]$. For each variable $x_i$, let $\alpha_i$ be the maximum
exponent of $x_i$ appearing among the monomials in $M$. The least common multiple
of the elements of $M$ is then
$\textstyle{\lcm(M) = \prod_{i=1}^n x_i^{\alpha_i}.}$

\begin{definition}
Let $I \subseteq \kb[\xb]$ be a monomial ideal with minimal monomial generating
set $G(I)=\{m_1,\dots,m_r\}$.  
The \emph{$\lcm$-lattice} of $I$ is the set
\[
L_I = \big\{ \lcm(\{m_i : i \in \sigma\}) \;\big|\; \sigma \subseteq \{1,\dots,r\} \big\},
\]
partially ordered by divisibility.
This poset is a finite atomic lattice, with atoms corresponding to the minimal
generators of $I$.
\end{definition}

The $\lcm$-lattice contains all possible interaction patterns among the minimal
generators of $I$. In particular, distinct subsets of generators may produce
the same least common multiple, and such coincidences reflect algebraic
redundancy caused by overlap among generators.
From this perspective, the size and shape of the $\lcm$-lattice provide a global
measure of how strongly the generators of $I$ interact. When all subsets of
generators yield distinct least common multiples, the $\lcm$-lattice is maximal
and coincides with the Boolean lattice. In contrast, extensive overlap among
generators forces many subsets to collapse to the same element of $L_I$,
resulting in a significantly smaller lattice.

The $\lcm$-lattice has been extensively studied in the context of free resolutions
of monomial ideals and their connections to atomic lattices; see
\cite{GPV99, MS05, P06, M13, C22} and the references therein.

While the $\lcm$-lattice captures all interactions among generators at once, it
is often useful to study how these interactions accumulate as larger collections
of generators are considered. This leads naturally to filtrations of monomial
ideals that reflect successive levels of interaction, providing a dynamic view
of the $\lcm$-structure \cite{MSW25}.







\section{The Erd{\H{o}}s-R{\'e}nyi random network model}
\label{sec:ER}
The Erd{\H{o}}s--R{\'e}nyi--Gilbert model is a simple and rich model for generating random graphs. 
We review its classical asymptotic properties in detail here, as they provide the direct mathematical blueprint for the algebraic phase transitions we establish in Section 4.
In its most common formulation introduced by Gilbert, an Erd{\H{o}}s--R{\'e}nyi graph $ER(n,p)$
is constructed with $n$ vertices (also called nodes in the network literature),
where each of the $\binom{n}{2}$ possible edges is independently added with
probability $p$; see~\citep{ER59, Gilbert, N18}.

\subsection{Asymptotic Properties}

As the number of vertices $n$ tends to infinity, the structural properties of an
Erd{\H{o}}s--R{\'e}nyi graph depend primarily on the average degree
$\langle k \rangle = (n-1)p \approx np$. We briefly recall several classical
asymptotic properties relevant to our discussion.

\subsubsection*{Degree distribution}
For large $n$ with $\langle k \rangle$ fixed, the degree distribution converges
to a Poisson distribution \citep{N18}:
\begin{equation}
    P(k) = e^{-\langle k \rangle} \frac{\langle k \rangle^k}{k!}.
\end{equation}
Thus, most vertices have degree close to the average, while vertices of very
high degree are exponentially rare \citep{BA16}.

\subsubsection*{Clustering coefficient}
The local clustering coefficient is defined as $C = p = \langle k \rangle / n$.
Consequently, for fixed $\langle k \rangle$ and $n \to \infty$, we have
$C \to 0$. This reflects the locally sparse nature of Erd{\H{o}}s--R{\'e}nyi
graphs and their lack of the high clustering typically observed in real-world
networks \citep{N18}.

\subsubsection*{Small-world property}
Despite low clustering, the average path length $L$ grows only logarithmically
with the number of vertices:
\begin{equation}
    L \sim \frac{\ln n}{\ln \langle k \rangle}.
\end{equation}
This logarithmic scaling is known as the \emph{small-world}
effect~\citep{BA16}.

\subsection{Phase Transition: The Emergence of the Giant Component}
A particularly important property of the Erd{\H{o}}s--R{\'e}nyi model is the abrupt change in its
global connectivity as the edge probability $p$ increases. This phenomenon is a
continuous (second-order) phase transition, commonly referred to as the
\emph{percolation transition}, first described in the seminal work of \citet{ER60}. It is reflected
in the asymptotic behavior of the size $N_G$ of the largest connected component
relative to the total number of vertices $n$ \citep{ER60, B01}.

This transition marks the onset of global connectivity from purely local
interactions, a phenomenon that will reappear in an algebraic form in the
setting of random monomial ideals.

\begin{itemize}
  \item{\bf Subcritical Phase ($\langle k \rangle < 1$):}
The graph consists of many small, disconnected components. The largest connected
component has logarithmic size, $N_G = O(\log n)$, and the graph is asymptotically
a forest of small trees.

\item{\bf Critical Point ($\langle k \rangle = 1$):} At the threshold of the phase transition, the largest connected component grows
rapidly and has size $N_G \sim n^{2/3}$.

    
\item{\bf Supercritical Phase ($\langle k \rangle > 1$):} A unique giant connected component (GCC) emerges, occupying a positive fraction
of the vertices, $N_G \sim S\cdot n$ with $S>0$. All remaining components are
microscopic, of size $O(\log n)$.

\end{itemize}

\begin{remark}
It is important to distinguish the existence of a giant component from  full connectivity of the graph. The network becomes fully connected (no isolated nodes remaining) only when:
\begin{equation}
    p > \frac{\ln n}{n}.
\end{equation}
This threshold is much higher than that of the percolation transition ($p > 1/n$); see~\citep{ER59, B01}.
\end{remark}

\section{Algebraic approach to the Erd{\H{o}}s-R{\'e}nyi model}
\label{sec:algER} 
\subsection{Combinatorial and algebraic density in $\Uk_{(1,n,2)}$}\label{sec:algER1}
The structural set of monomials $\Uk_{(1,n,2)}$ is in one-to-one correspondence
with the set of all simple graphs on $n$ vertices, $\Gk_n$. To each graph
$G=(V,E)$ in $\Gk_n$ we assign its corresponding monomial ideal
$I_G=\langle x_i x_j \mid (i,j)\in E\rangle.$
Conversely, every squarefree monomial ideal in $n$ variables that is
equigenerated in degree $2$ is the monomial edge ideal of some graph in $\Gk_n$.
In this case, the combinatorial density of a set
$\Sc\subseteq\Uk_{(1,n,2)}$ coincides with the usual graph density of the
corresponding graph $G=(V,\Sc)$, which we denote by $gd(\Sc)$ or $gd(G)$.

\begin{figure}[h]
\centering
\includegraphics[width=0.9\linewidth]{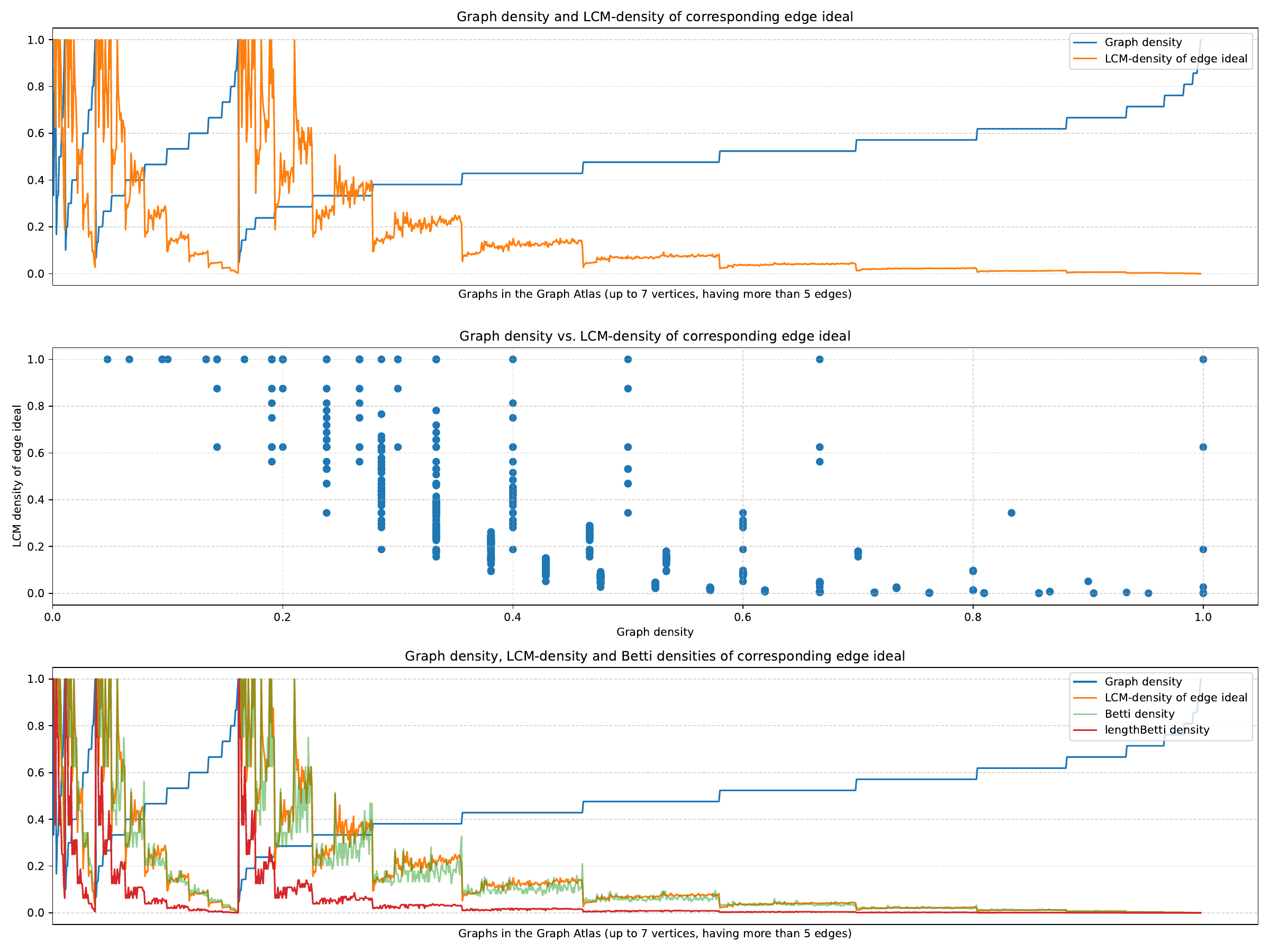}
\caption{Graph density, lcm density, Betti density and length Betti density  for all graphs in at most $7$ vertices.}
\label{fig:atlas_gden_lcmden}
\end{figure}

The Atlas of Graphs \cite{RW98} proposes a systematic description of the
collection of all graphs with $n$ vertices, for each $n$. The \texttt{Python}
library \texttt{networkx} \cite{networkx} implements this atlas via the function
\texttt{nx.graph\_atlas}, which lists all graphs with at most $7$ vertices.
Figure~\ref{fig:atlas_gden_lcmden} plots the relationship between the density of
each graph with at most $7$ vertices and the algebraic (or poset) density, the
Betti density, and the length-Betti density of the corresponding edge ideal.
We observe a negative correlation between graph density and the $\lcm$-density
of the edge ideal; see Table~\ref{tab:corr_matrix_upto7}. This correlation is
more pronounced when restricting to graphs with $7$ vertices; see
Table~\ref{tab:corr_matrix_7}. The tables also show a high correlation between
$\lcm$-density and Betti density.

\begin{figure}[htp]
\subfloat[Density and graph density for all graphs in $6$ vertices with no isolated vertices.]{%
\includegraphics[width=0.6\linewidth, angle=270]{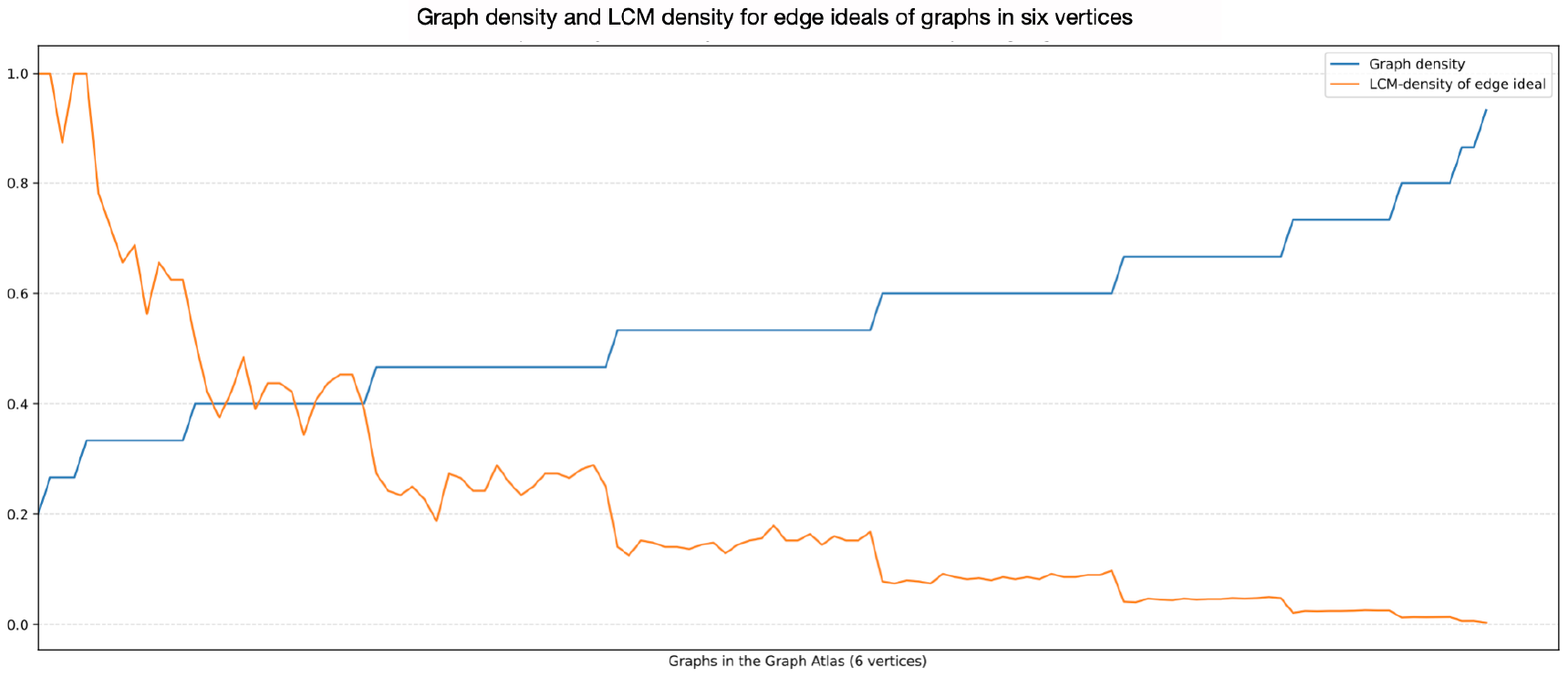}
}

\subfloat[Density and graph density for all graphs in $7$ vertices with no isolated vertices.]{%
\includegraphics[width=0.9\linewidth]{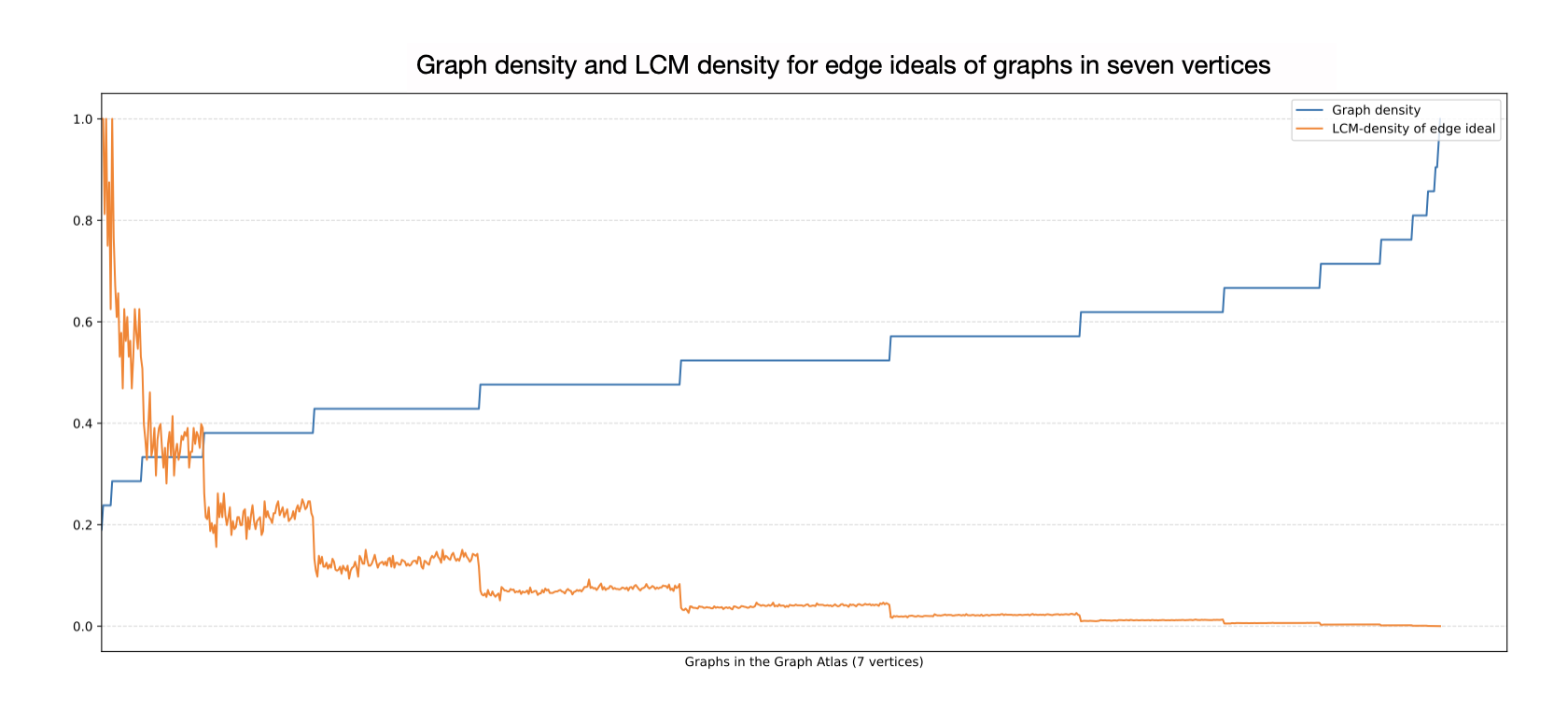}
}
\caption{Density and graph density for all graphs in $6$ and $7$ vertices with no isolated vertices.}
\label{fig:atlas_gden_lcmden_6-7}
\end{figure}

\begin{table}[htp]
\begin{center}
\begin{minipage}[t]{0.45\textwidth}%
\begin{tabular}{ccccc}
\toprule
 & gd & d & bd & lbd \\
\midrule
gd & 1.0000 & -0.6902 & -0.6535 & -0.4721 \\
d & -0.6902 & 1.0000 & 0.9794 & 0.8439 \\
bd & -0.6535 & 0.9794 & 1.0000 & 0.8695 \\
lbd & -0.4721 & 0.8439 & 0.8695 & 1.0000 \\
\bottomrule
\end{tabular}
\caption{Pearson correlation matrix for densities of graphs with up to $7$
vertices, where $gd$ denotes graph density, $d$ the $\lcm$-density, $bd$ the
Betti density, and $lbd$ the length-Betti density.}
\label{tab:corr_matrix_upto7}
\end{minipage}
\hfill
\begin{minipage}[t]{0.45\textwidth}%
\begin{tabular}{ccccc}
\toprule
 & gd7 & d7 & bd7 & lbd7 \\
\midrule
gd7 & 1.0000 & -0.7903 & -0.7678 & -0.5912 \\
d7 & -0.7903 & 1.0000 & 0.9749 & 0.8360 \\
bd7 & -0.7678 & 0.9749 & 1.0000 & 0.8627 \\
lbd7 & -0.5912 & 0.8360 & 0.8627 & 1.0000 \\
\bottomrule
\end{tabular}
\caption{Pearson correlation matrix for densities of graphs with $7$
vertices, where $gd7$, $d7$, $bd7$, and $lbd7$ denote the corresponding densities
restricted to graphs on $7$ vertices.}
\label{tab:corr_matrix_7}
\end{minipage}
\end{center}
\end{table}

The data shown in Figures~\ref{fig:atlas_gden_lcmden} and
\ref{fig:atlas_gden_lcmden_6-7} reveal, for each $n$, three distinct regimes.
In the low graph density regime, the $\lcm$-density of the corresponding edge
ideal is higher than the graph density and exhibits substantial variability.
This is followed by a transition regime (graph density $gd=0.4$ for graphs on
$6$ vertices and $gd=10/3$ for graphs on $7$ vertices), in which some graphs have
$\lcm$-density higher than their graph density, while others have it lower.
Finally, in the high graph density regime, the $\lcm$-density of the edge ideal
shows a strong negative correlation with graph density and decreases toward
zero.

Motivated by these observations, we formulate the following conjecture.
\begin{conjecture}\label{conj:densityPhaseTransition}
The $lcm$-density of edge ideals of graphs exhibits a phase transition phenomenon
when considering the set of all graphs on $n$ vertices as $n\to\infty$.
\end{conjecture}


The Erd{\H{o}}s--R{\'e}nyi model applies verbatim to $\Uk_{(1,n,2)}$, and
Theorem~3.10 in \cite{MSW25} shows that
Conjecture~\ref{conj:densityPhaseTransition} holds for the family of
$ER(n,p)$-model graphs (or, equivalently, the corresponding monomial ideals).

\subsection{$\Uk_{(1,n,d)}$: equigenerated monomial ideals of degree $d> 2$}\label{sec:algER2}
The Erd{\H{o}}s--R{\'e}nyi model can be directly generalized to
$\Uk_{(1,n,d)}$ for $d>2$, and many of its properties admit natural analogues.
The monomial ideals in this set correspond to edge ideals of $d$-uniform
hypergraphs.

An Erd{\H{o}}s--R{\'e}nyi model for equigenerated squarefree monomial ideals of
degree $d$ is defined as follows. Start with an empty collection $\Sc$ of
$d$-subsets and fix a probability $0<p<1$. For each element of
$\Uk_{(1,n,d)}$, that is, for each $d$-subset
$\sigma\subseteq\{1,\dots,n\}$, include $\sigma$ in $\Sc$ independently with
probability $p$. The expected size of $\Sc$ is
$p\cdot\binom{n}{d}$. If we define the density of a collection $\Sc$ by
\[
\cd(\Sc)=\frac{|\Sc|}{\binom{n}{d}},
\]
then the expected density of $\Sc$ is $p$.

For any $d$-subset $\sigma\subseteq\{1,\dots,n\}$, define the monomial
$\xb_\sigma=\prod_{i\in\sigma}x_i$, which has degree $d$. The monomial ideal
$I_\Sc=\langle \xb_\sigma \mid \sigma\in\Sc\rangle$
is a squarefree monomial ideal equigenerated in degree $d$, and every such
ideal arises in this way. Hence, this construction defines an Erd{\H{o}}s--R{\'e}nyi
model for squarefree equigenerated monomial ideals. Throughout, we restrict
without loss of generality to collections $\Sc$ with full support, that is,
$\bigcup_{\sigma\in\Sc}\sigma=\{1,\dots,n\}$.

Our goal in this section is to study the behavior of the $\lcm$-density of
squarefree equigenerated monomial ideals, hence generalizing the case of edge
ideals. Our two guiding statements are:
\begin{itemize}
\item There is a phase transition for the $\lcm$-density of equigenerated
squarefree monomial ideals.
\item The logarithm of the $\lcm$-density of equigenerated squarefree monomial
ideals exhibits a strong negative correlation with the density of the
corresponding collection of generators.
\end{itemize}

\noindent To support these statements, we consider the following example.

\begin{figure}[htp]
\includegraphics[width=\linewidth]{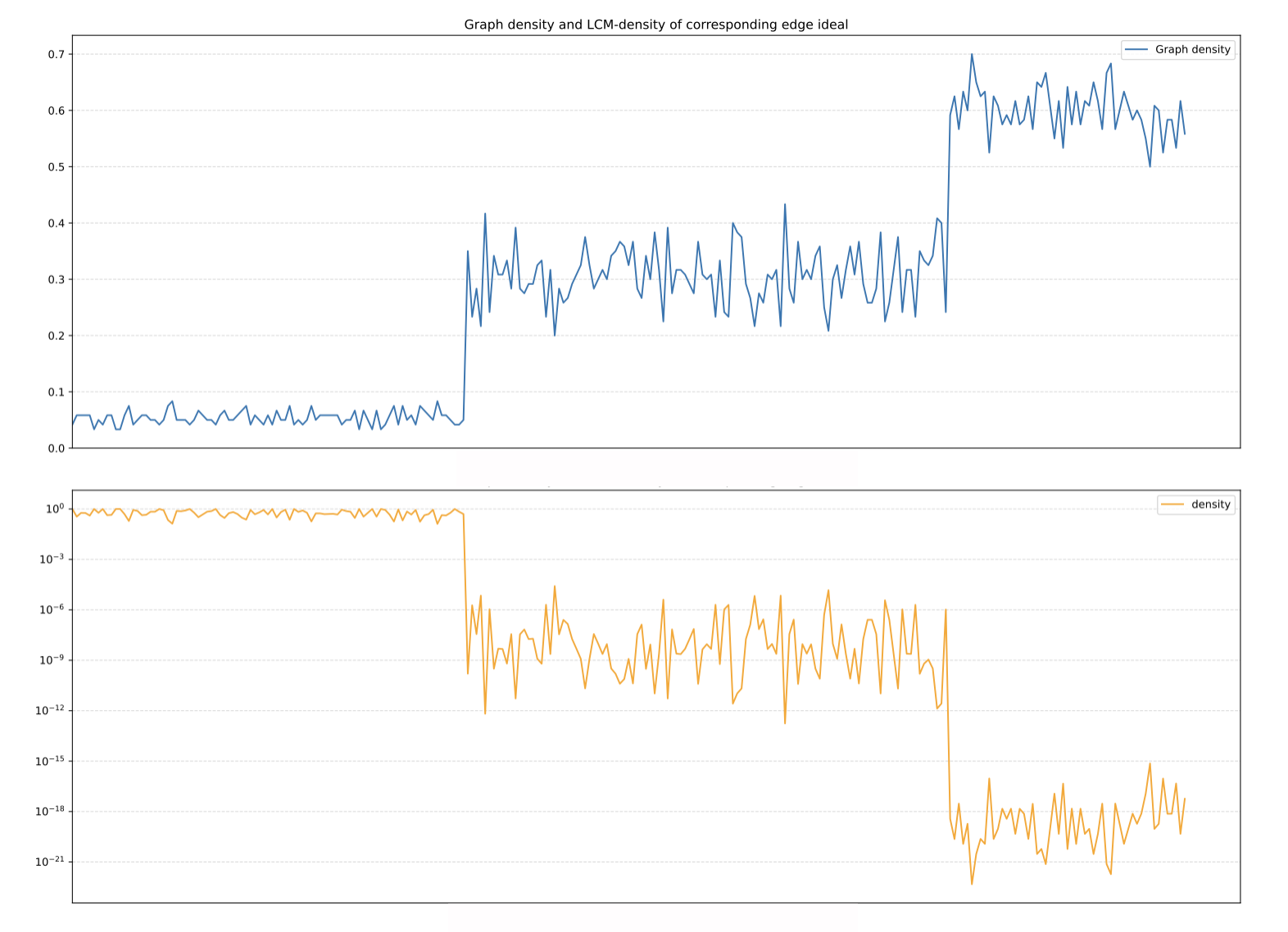}
\caption{Collection density and $\lcm$-density for samples from the
$ER(10,k,p)$ model with $n=10$, $k=3$, and $p\in\{0.03,0.3,0.6\}$. The
$\lcm$-densities are plotted on a logarithmic scale.}
\label{fig:3-out-of-10_correlation}
\end{figure}

\begin{figure}[htp]
\begin{center}
\includegraphics[width=0.9\linewidth]{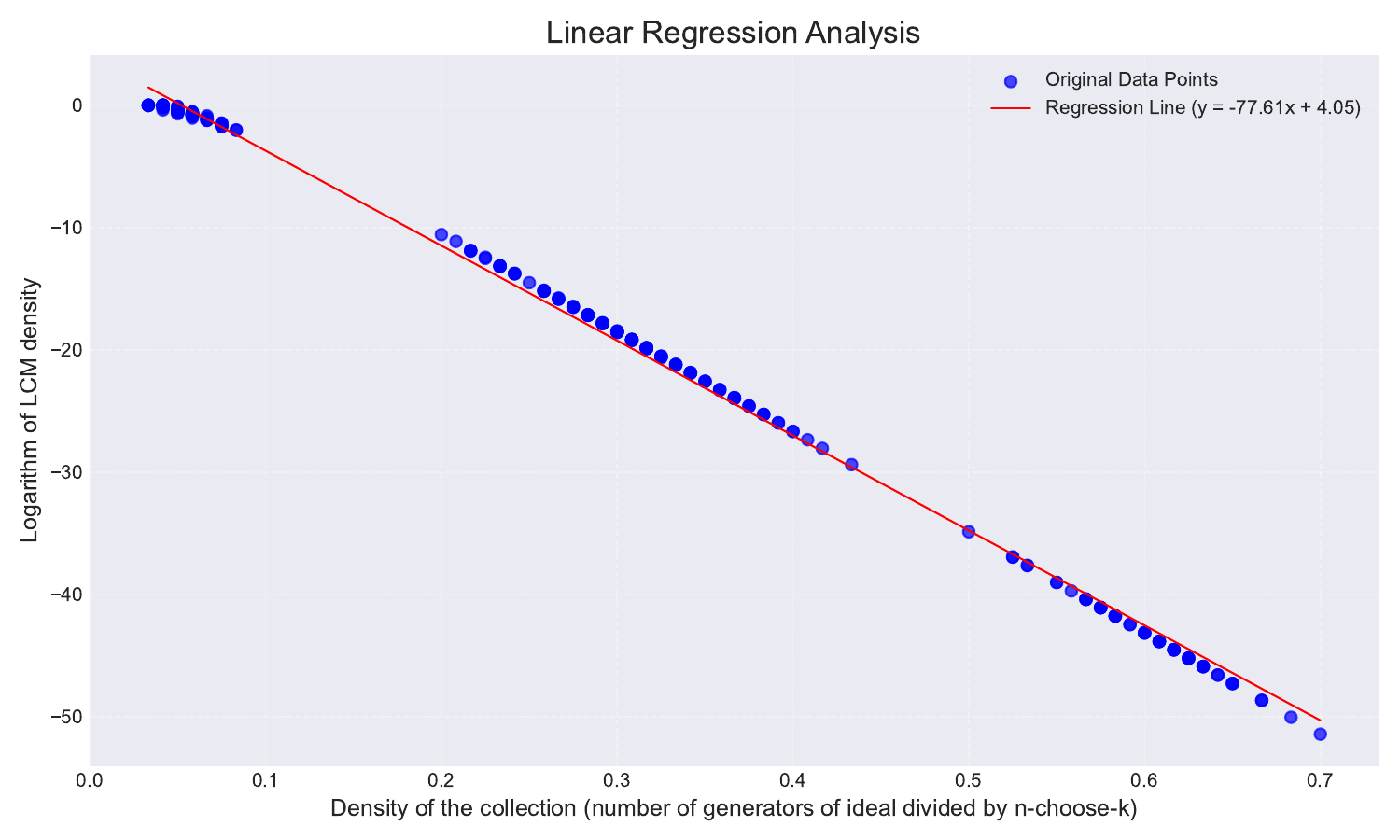}
\caption{Linear regression plot between collection density and logarithm of $\lcm$-density.}
\label{fig:regression_3_10}
\end{center}
\end{figure}

\begin{example}
Let $n=10$ and $d=3$, and consider the probabilities $p=0.03$, $p=0.3$, and
$p=0.6$. For each value of $p$, we generate a random sample of
$\frac{100}{dp}$ collections and retain only those with full support. For each
such collection $\Sc$, we compute the corresponding monomial ideal $I_\Sc$ and
its algebraic density.

Figure~\ref{fig:3-out-of-10_correlation} shows the density of each collection of
$d$-subsets together with the $\lcm$-density of the corresponding ideal. The
$\lcm$-densities are plotted on a logarithmic scale, while the collection
densities are plotted on a linear scale. The plot shows an almost perfect
inverse relationship between these two quantities. This behavior is quantified
in Table~\ref{tab:3-out-of-10_correlation}, which shows an extremely high
correlation. A corresponding regression line is displayed in
Figure~\ref{fig:regression_3_10}.

The regression indicates the existence of $\alpha,\beta\in\RR$ such that
\[
\log\!\left(\frac{|L_{I_\Sc}|}{2^{r}}\right)
=
\alpha\,\frac{r}{\binom{n}{d}}+\beta,
\]
equivalently,
\[
\log(|L_{I_\Sc}|)
=
r\!\left(1+\frac{\alpha}{\binom{n}{d}}\right)+\beta.
\]
This yields
\[
|L_{I_\Sc}|
=
2^{\,r\left(1+\frac{\alpha}{\binom{n}{d}}\right)+\beta},
\]
showing that the size of the $\lcm$-lattice of equigenerated squarefree monomial
ideals of degree $d$ depends only on the number of generators.
\end{example}

\begin{table}[htp]
\begin{center}
\caption{Pearson correlation matrix}
\label{tab:3-out-of-10_correlation}
\begin{tabular}{ccc}
\toprule
 & collection density & log of \lcm-density \\
\midrule
collection density & 1.000000 & -0.999093 \\
log of \lcm-density & -0.999093 & 1.000000 \\
\bottomrule
\end{tabular}
\end{center}
\end{table}

\section{General random models and parametric families of ideals }
The  Erd{\H{o}}s--R{\'e}nyi model for squarefree equigenerated monomial ideals can be seen as a special case of the random monomial ideal model introduced in \cite{RMI}, which was also inspired by the study of random graphs and simplicial complexes. This broader class of probabilistic models for random monomial ideals, not necessarily squarefree, is defined as follows. 
Fix an integer $D$ and a parameter $p=p(n,D)$, $0\leq p\leq 1$. Construct a random set of monomial ideal generators $B$ by including, independently, with probability $p$ each non-constant monomial of total degree at most $D$ in $n$ variables. The resulting random monomial ideal is simply $I=\langle {B}\rangle$, and if $B=\emptyset$, then let $I=\langle{0}\rangle$. 
The notation for this data-generating process uses $\mathcal B(n,D,p)$ to denote the resulting distribution on the sets of monomials. This distribution on monomial sets induces a distribution on the set of ideals, called the \er-type distribution on monomial ideals and denoted by $\mathcal{I}(n,D,p)$. 

It is important to note that the set $\mathcal{I}(n,D,p)$ is a parametric family of probability distributions, with one distribution for each choice of the parameters $n$, $D$, and $p$. 
In \cite[Section~5]{RMI}, the authors define the \emph{graded} model, which allows the probability parameter $p$ to be a vector of length $D$, so that each degree $i$ of a minimal generator is associated with a probability $p_i$. 
If in the graded model there is only one non-zero entry $p_d$ in the parameter vector $p$, then the ideals are generated in degree $d$. 
The behavior of general random monomial ideals appears to mimic that of graphs and squarefree ideals. 

Equigenerated random monomial ideals were studied in \cite{Average_Min_Res}, where the authors study phase transition phenomena for genericity and Scarfness. 
These properties are related to lcm-lattice density but not equivalent; ideals whose Taylor resolutions are minimal--and whose $\lcm$-lattices are therefore of maximal size--are examples of Scarf ideals. The main results of~\cite{Average_Min_Res} are asymptotic, and consider a regime in which the degree $D \to \infty$. In this setting, when the  probability parameter $p_D$ is sufficiently large as a function of $D$, randomly generated monomial ideals are almost never Scarf. 
In general, there are two asymoptotic regimes; the second one fixing $D$ and letting $p=p(n)$ vary with the number of variables, as $n\to\infty$. 

In equigenerated \er-type random monomial ideals, every generator is minimal. Thus the number of generators is directly correlated with the probability parameter $p_d$ for selecting degree-$d$ monomials (contrast this with \cite[Theorem 4.1]{RMI}).

\begin{figure}[htp]
\begin{center}
\includegraphics[width=0.45\linewidth]{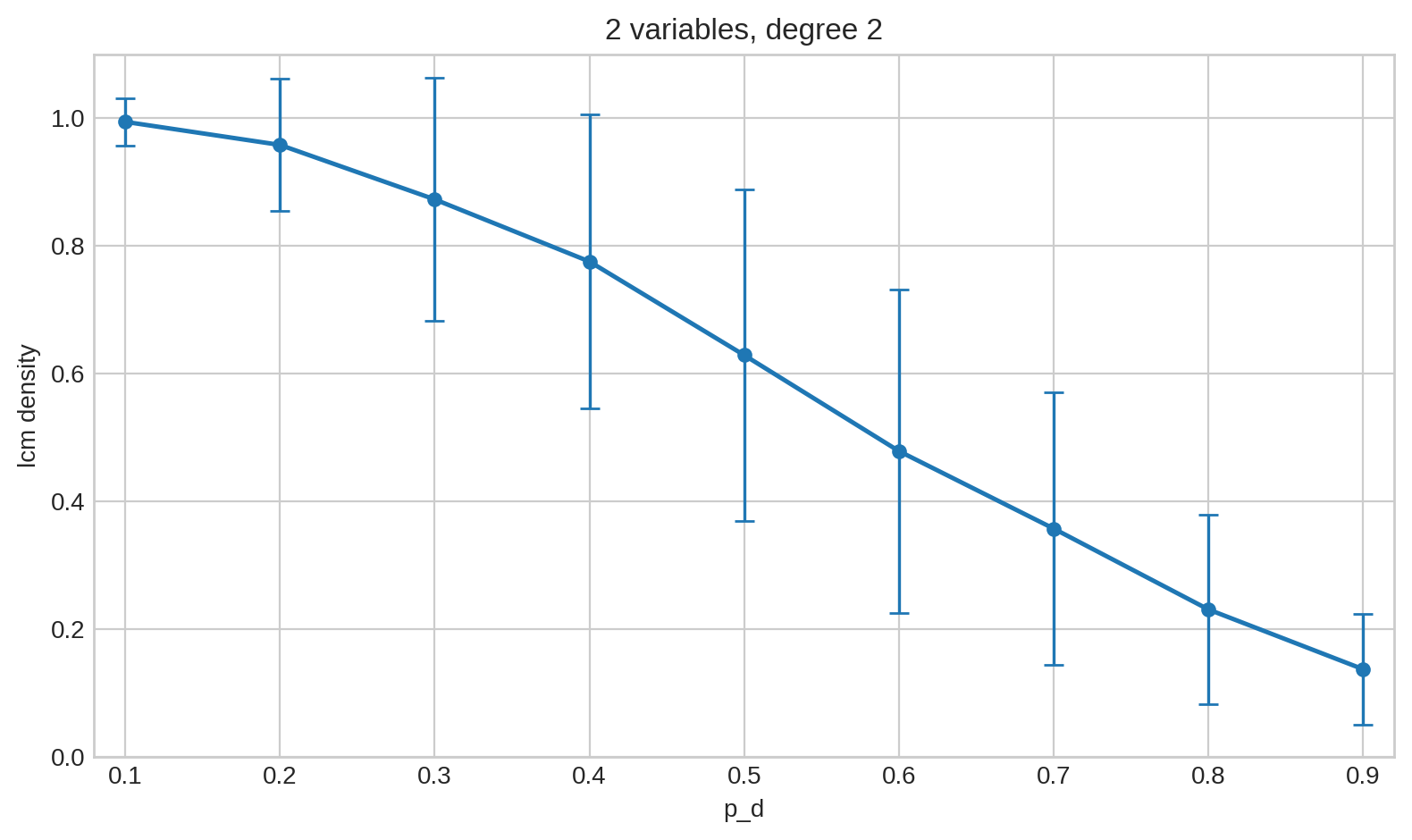}
\includegraphics[width=0.45\linewidth]{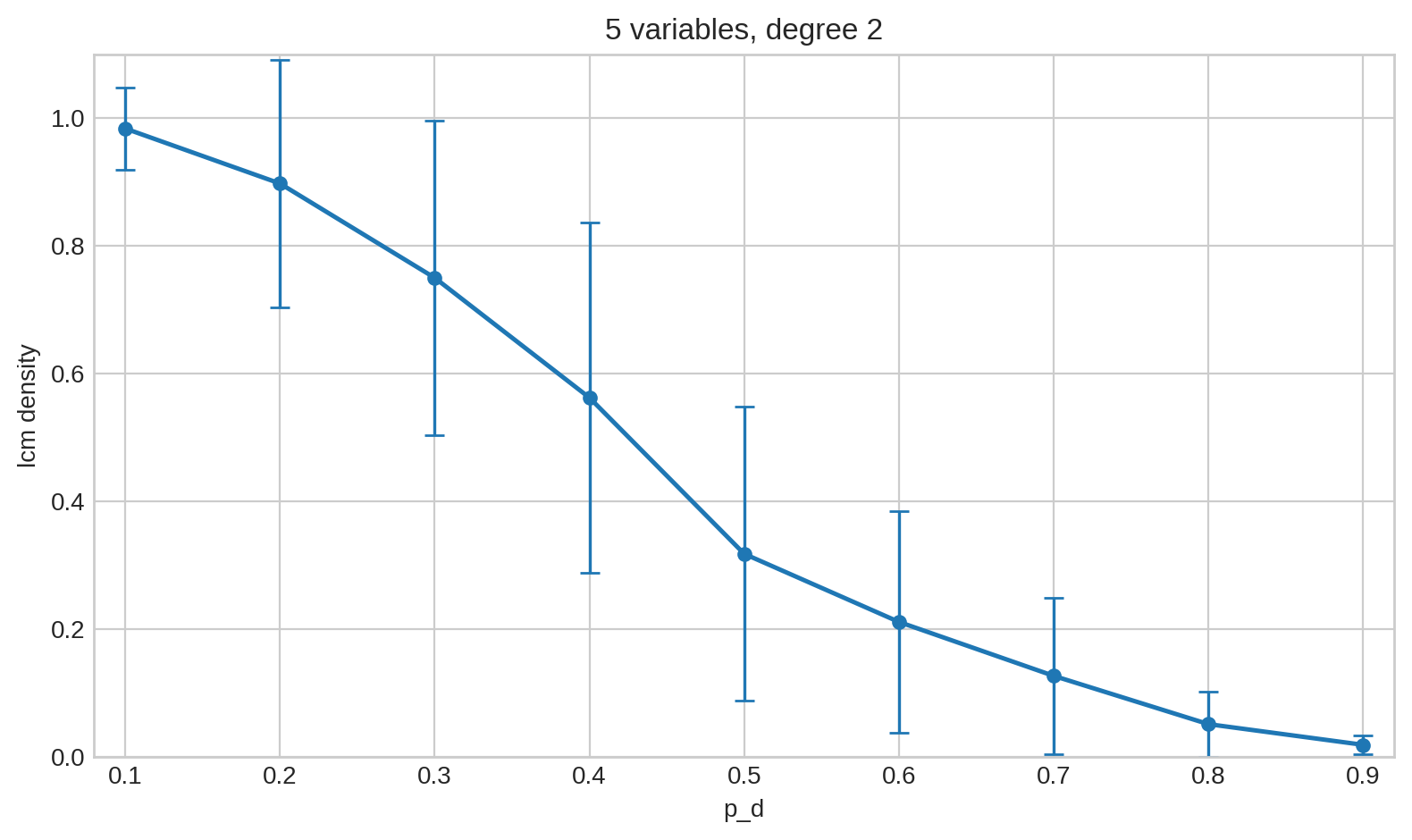}
\caption{Mean and standard deviation of $\lcm$-density for equigenerated random monomial ideals in degree $d=2$ in $n=2$ variables (left) and $n=5$ variables (right). For each value of $n,d,p$, the sample contains $100$ monomial ideals.}
\label{fig:MeanStdev density for RMI}
\end{center}
\end{figure}

If one interprets the squarefree condition as a sampling restriction on samples from the $\mathcal I(n,D,p)$ model, the results here further support our previous observation: that the size of the $\lcm$-lattice of equigenerated monomial ideals (squarefree or not) of degree $d$ depends heavily -- but not only -- on the number of generators. One possible difference in this model is that we do not only retain monomials with full support. 
However, the \emph{mean} $\lcm$-density of a randomly generated monomial ideal in degree $d$ appears to depend only on the number of generators as well.

These results indicate two things:
\begin{itemize}
    \item The \er-type random model for monomial ideals captures the $\lcm$-density behavior. 
    \item The sample mean of $\lcm$-density of equigenerated random monomial ideals is a good predictor for the behavior shown in  Figure~\ref{fig:3-out-of-10_correlation}, for smaller number of variables $n$. 
\end{itemize} 
\begin{figure}[htp]
\begin{center}
\includegraphics[width=0.9\linewidth]{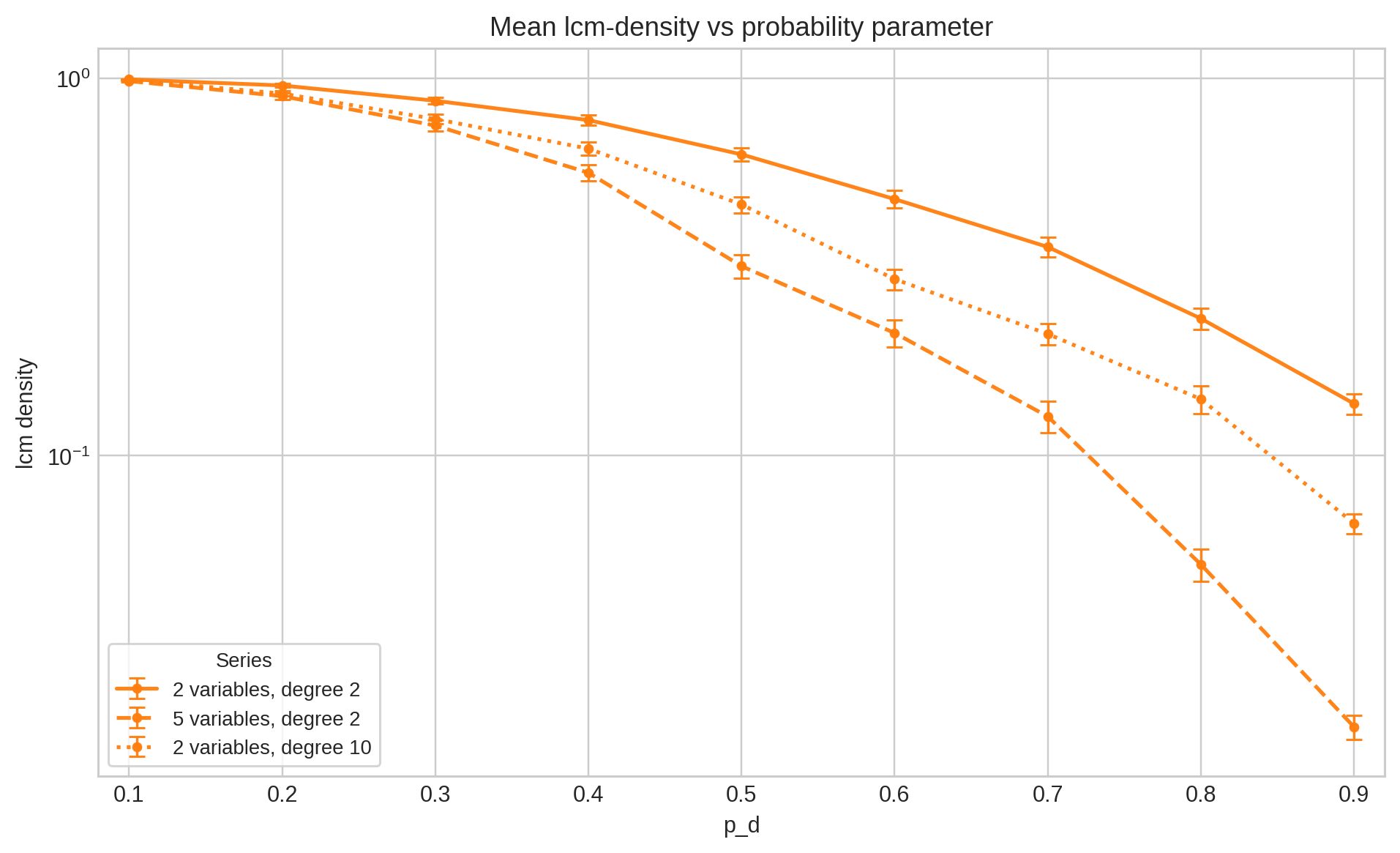}
\caption{Mean and its standard error for $\lcm$-density for equigenerated random monomial ideals generated from the model $\mathcal I(n,d,p)$. For each value of $n,d,p$, the sample contains $100$ monomial ideals. The $\lcm$-density is shown on a log scale.}
\label{fig:Mean is a good estimator for lcm density using RMI}
\end{center}
\end{figure}

Given that generating large samples of monomial ideals on many variables is non-trivial, we showcase the behavior of the mean $\lcm$-densities for small $n$ and varying degrees $d$ for the three values of probability parameter that match Figure~\ref{fig:3-out-of-10_correlation}. 

\begin{figure}[htp]
\begin{center}
\includegraphics[width=0.9\linewidth]{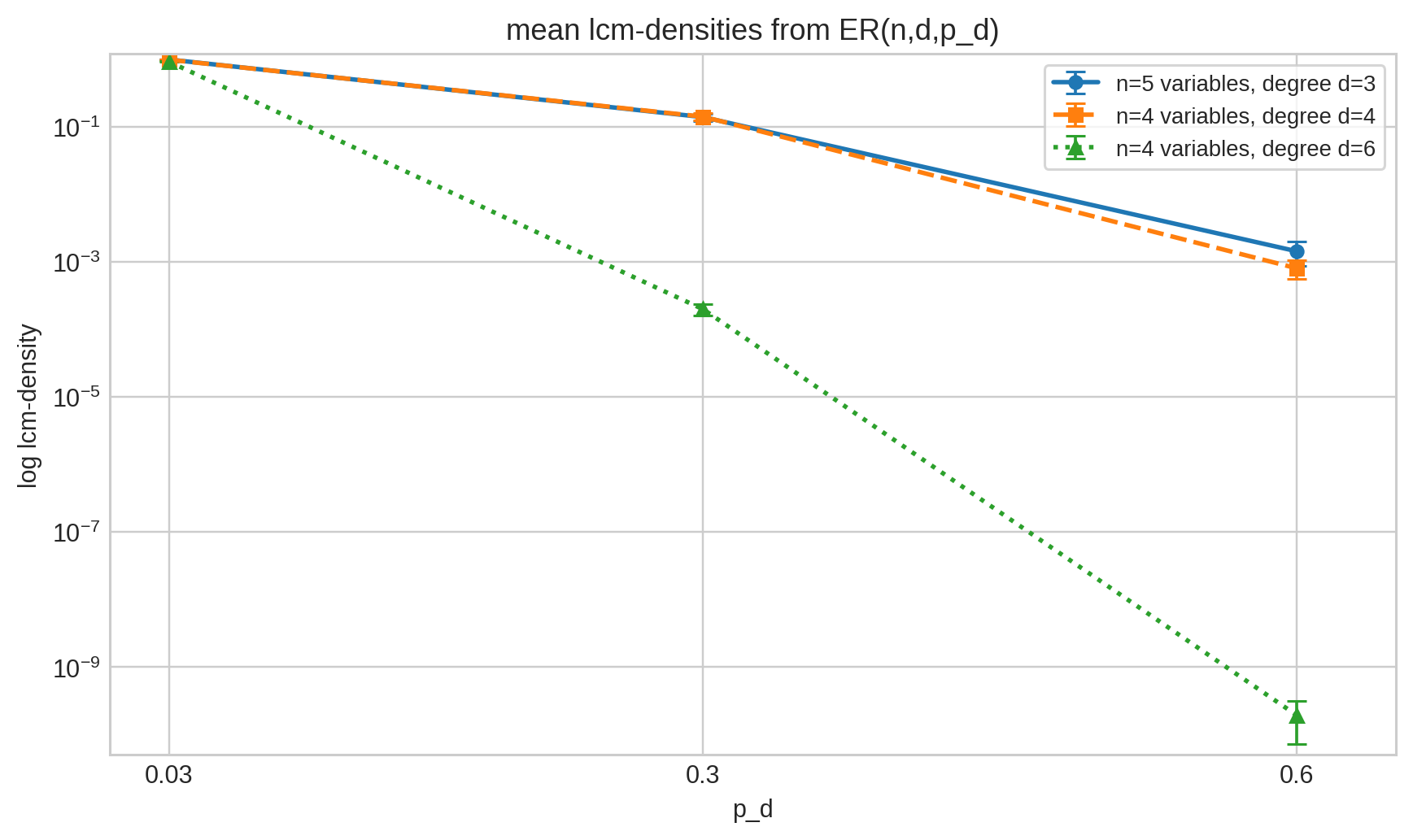}
\caption{Mean and its standard error for $\lcm$-density for equigenerated random monomial ideals generated from the model $\mathcal I(n,d,p)$, for $p\in\{0.03,0.3,0.6\}$.  For each value of $n,d,p$, the sample contains $100$ monomial ideals.}
\label{fig:Mean is a good estimator for lcm density using RMI}
\end{center}
\end{figure}

\noindent{\bf Code Availability.}
All computational experiments in this paper were carried out using code that is publicly available at\\
https://github.com/FatemehMohammadi/AsymptoticPropertiesRandomIdeals.
\begin{acks}
This work is partially supported by grant PID2024-157733NBI00 funded by MCIN/AEI/10.13039/501100011033/FEDER EU, the FWO grants G0F5921N (Odysseus), G023721N,  the KU Leuven grant iBOF/23/064, and the Simons Foundation Travel Support for Mathematicians Gift 854770.  
\end{acks}

\bibliographystyle{ACM-Reference-Format}
\bibliography{MPS_bibliography}


\end{document}